\documentclass[aps, twocolumn, nofootinbib, showpacs, prd]{revtex4}
\usepackage{graphicx}
\usepackage{amsfonts}
\usepackage{amssymb}
\usepackage{amsbsy}
\usepackage{amsmath}
\usepackage{latexsym}
\usepackage{natbib}
\usepackage{bm}
\usepackage{color}
\usepackage{float}

\begin{document}
\title{Geometric Selection Rules for Singularity Formation in Modified Gravity}

\author{Soumya Chakrabarti}
\email{soumya.chakrabarti@vit.ac.in}

\affiliation{School of Advanced Sciences, Vellore Institute of Technology, Vellore, Tiruvalam Rd, Katpadi, Tamil Nadu 632014, India}


\date{\today}

\begin{abstract}
We argue that the polynomial degeneracies of curvature invariants can act as geometric selection rules for spacetime singularities in modified theories of gravity. The degeneracies arise purely from the algebraic structure of Riemannian geometry and impose non-trivial constraints on the effective energy-momentum tensor. We derive these constraints for metric $f(R)$ gravity and a wide class of scalar-tensor theories to show that a singularity formation is generally occluded by curvature and/or scalar-induced anisotropies. Therefore, formation of a singularity in modified theories of gravity is not always a generic outcome but can occur only along algebraically admissible branches selected by Riemannian curvature invariants.
\end{abstract}

\maketitle

\section{Introduction}
Spacetime singularities remain one of the most fundamental yet less understood concepts in gravitational physics. They arise as limiting configurations of General Relativity (GR) in which curvature invariants diverge and the classical description of spacetime loses its predictability, leading to geodesic incompleteness \cite{penrose0, penrose00}. The Schwarzschild spacetime represents a canonical example : a static, spherically symmetric vacuum geometry containing a curvature singularity at the vanishing areal radius \cite{sch}. In GR, singularities are either hidden behind trapped surfaces (black holes) \cite{bh1} or visible to distant observers (naked singularities) \cite{penrosens, ns1, ns2}. Gravitational collapse remains the only known natural astrophysical process through which massive objects are dynamically driven towards such strong curvature regime traditionally associated with singular limits. However, it is important to note that a collapse need not necessarily culminate in a singularity, as well-motivated non-singular end states do exist, such as regular black holes and wormholes, for which curvature invariants remain finite everywhere \cite{regular, sc}. Despite extensive investigations and various formulations of cosmic censorship conjectures \cite{censor1, censor2}, no general principle is known that determines, solely from initial data, whether gravitational evolution must terminate in a singularity.  \\

This ambiguity becomes even more pronounced in modified theories of gravity. Extensions of GR, such as theories $f(R)$ or scalar-tensor theories, introduce additional scalar degrees of freedom and modify the field equations to accommodate an effective stress-energy content for spacetime \cite{fR, st}. There are claims in the literature that spacetime singularities can be weakened or their formation completely avoided in modified theories of gravity, typically grounded in the construction of analytical/numerical solutions of these field equations \cite{nonsing}. Although such approaches can indeed produce valuable insights, they are often technically demanding and strongly model-dependent. Additionally, collapsing solutions in modified theories require a consistent imposition of junction conditions beyond GR. While in GR, a continuity of the induced metric and extrinsic curvature on the boundary hypersurface is sufficient \cite{GRjunction}, modified gravity theories usually demand an additional matching of curvature invariants such as the Ricci scalar and its' spatial derivative across the boundary hypersurface \cite{junction}. These additional matching conditions severely restrict the construction of a consistent set of interior-exterior solutions. As a consequence, deductions related to singularities in modified gravity are usually tied to particular symmetry assumptions rather than a generic formalism. This motivates us to ask a foundational question : \emph{beyond the field equations or matching prescriptions, do purely geometric constraints derived from Riemannian geometry naturally restrict the admissibility of spacetime singularities in modified theories of gravity?}

We address this question in this article in the following order : Section $II$ includes a summary of polynomial degeneracies of Riemannian geometry and the intuitions that can be built from them in a theory-agnostic manner. Section $III$ and $IV$ extends these intuition into specific modified theories, namely, $f(R)$ gravity and a scalar-tensor theory. In Section $V$ and $VI$ we explore the signature of these degeneracies on one of the most celebrated scalar extended theory of gravity, the Brans-Dicke theory and its generalization. The additional purpose of these two sections is to see if it is possible to find a geometric bound on the Brans-Dicke parameter. In Section $VII$ we consider a modified theory of gravity where the scalar field interacts with ordinary matter energy-momentum distribution and comment on the constraints on the interaction profiles. We form a correlaton of polynomial degeneracy with geodesic deviation for a family of null geodesics using the Raychaudhuri equation, in Section $VIII$. We conclude the article in Section $IX$.

\section{Polynomial Degeneracy of Riemannian Geometry : Framework}
A natural framework to address this question can be found in the polynomial degeneracies of curvature invariants in Riemannian geometry. The primary idea of a so-called 'degeneracy' is simple : \emph{not all scalars constructed from contractions of the Riemann tensor are algebraically independent}. Even when functionally independent, these scalars satisfy non-trivial polynomial identities, known as syzygies \cite{syzygy0, syzygy1}. Identifying a complete set of independent Riemannian invariants is a challenging mathematical task, owing to the complexity of solving the associated degeneracy relations \cite{syzy11, syzy111, syzy2, syzygy22}. In the course of analyzing the spacetime dynamics near a curvature singularity, usually strong simplifying assumptions are imposed at the outset, such as Ricci flatness or highly symmetric matter sources. While such assumptions do facilitate a construction of exact solution, they drastically reduce the number of admissible invariants \cite{syzy3}. In contrast, avoiding a premature simplification can always preserve the observer-independent character of curvature invariants and their gradient flows, allowing generic geometric constraints to be extracted. In this spirit, we work with a hierarchy of invariants constructed from the trace-free Ricci tensor and the Weyl tensor \cite{syzy4} without imposing any restrictions at the outset. The trace-free Ricci tensor is defined as
\begin{equation}\label{tracefree}
S^{\alpha}{}_{\beta} = R^{\alpha}{}_{\beta} - \frac{R}{4}\delta^{\alpha}{}_{\beta}.
\end{equation}
From this tensor one can construct the scalar invariants as
\begin{equation}\label{r1}
r_1 = S^{\alpha}{}_{\beta}S^{\beta}{}_{\alpha},~ r_2 = S^{\alpha}{}_{\beta}S^{\gamma}{}_{\alpha}S^{\beta}{}_{\gamma},~r_3 = S^{\alpha}{}_{\beta}S^{\gamma}{}_{\alpha}S^{\delta}{}_{\gamma}S^{\beta}{}_{\delta}.
\end{equation}

Using the corresponding field equations of the theory, these invariants may be expressed algebraically in terms of the energy-momentum tensor, for instance, as
\begin{equation}\label{r1t}
r_1 = \frac{1}{4} \left(T^{\alpha}{}_{\beta}T^{\beta}{}_{\alpha}-\frac{T^2}{4}\right),
\end{equation}
with analogous expressions for $r_2$ and $r_3$. In addition, one may also construct Weyl invariants as
\begin{equation}\label{w2}
w_{2} \sim C_{\alpha\beta\lambda\delta}C^{\alpha\beta\gamma\psi}C^{\lambda\delta}{}_{\gamma\psi}.
\end{equation}

We emphasize once more that these invariants are not algebraically independent. One can use a chosen set of field equations for a particular theory of gravity and translate the syzygies into algebraic restrictions on the effective matter sector \cite{syzy3, syzy4, syzy5}. While these identities are automatically satisfied for simple sources of matter, such as dust or perfect fluids, they impose nontrivial conditions for more general configurations involving an anisotropic stress or energy flux. The earliest study on these conditions by Narlikar and Karmarkar \cite{narlikar} have proved that in spherically symmetric spacetimes only four invariants, $\{R,r_1,r_2,w_1\}$, are truly independent. For more discussions related to classification of curvature invariants and their syzygies in warped product spacetimes (which encompass spherical, planar, and hyperbolic symmetry classes) we refer to the following articles \cite{koutras, tipler, syzy3, syzy4, syzy11, syzy111, carot, nakahara, haddow}. For a spherically symmetric spacetimes, the independent invariants satisfy the algebraic relation
\begin{equation}
(-12r_{3}+7r_{1}^{2})^{3} - (12r_{2}^{2}-36r_{1}r_{3}+17r_{1}^{3})^{2} = 0.
\label{r33}
\end{equation}

Higher-order Ricci invariants are subject to additional algebraic relations arising from further polynomial syzygies, which may be generated in a systematic manner, as reported in the literature up to $r_{10}$ \cite{syzy3}. An important and often overlooked feature of Ricci invariants is their observer independence: they are constructed solely from contractions of the Ricci tensor and do not involve any reference to a timelike four-velocity $u^{\alpha}$. While the energy-momentum tensor $T^{\alpha}{}_{\beta}$ may single out a preferred four-velocity, such as the fluid rest-frame in the case of a perfect fluid, this choice does not enter the definition of Ricci invariants. More general timelike observers therefore play no role in the algebraic structure of these scalars and they can encode information about the matter content in a frame-independent manner. This information can become crucial in the context of a gravitational collapse, where the nonlinear nature of gravity causes curvature to reinforce itself rapidly as collapse proceeds toward a sudden or an asymptotic divergence. For a spherically symmetric metric in particular, these algebraic identities among curvature invariants can play a decisive role in determining whether divergences are admissible. Reformulating the question in a different manner, we ask : \emph{can invariant geometric constraints alone determine whether singularities are admissible in modified gravity?}  \\

For a perfect fluid with energy-momentum tensor $T_{\alpha \beta} = (\rho + p) u_{\alpha} u_{\beta} + p g_{\alpha \beta}$, the Ricci invariants reduce to a particularly simple form
\begin{equation}\label{perfect}
r_{n} \sim (\rho + p)^{n+1}.
\end{equation}

For the perfect fluid, Eq. (\ref{r33}) is identically satisfied. The situation changes remarkably once the matter sector admits anisotropic stresses and heat flow. We keep in mind that the effective energy-momentum tensor for a generic modified gravity can be recast as an imperfect fluid \cite{goncamoss, scalarcollapse1, capo, scalarcollapse2, scalarcollapse3, nbsc, faracote, zimdahl, cho, giardino}, written as
\begin{equation}\label{EM_gen}
T_{\alpha \beta} = (\rho_{m} + p_t) u_{\alpha} u_{\beta} + (p_{r} - p_t) n_{\alpha} n_{\beta} + p_t g_{\alpha\beta} + 2 u_{(\alpha}q_{\beta)},
\end{equation}
with distinct radial and tangential pressures and a non-vanishing radial heat flux. Substituting Eq. (\ref{EM_gen}) into the syzygy (\ref{r33}) yields a nontrivial algebraic restriction,
\begin{equation}
\big \lbrace q_{\alpha}q^{\alpha}-(n_{\alpha}q^{\alpha})^{2}\big\rbrace^{2}(p_{r}-p_t)^{2} {\bf \it M} = 0.
\label{P}
\end{equation}
Here ${\bf \it M}$ denotes a polynomial function of the variables, namely, effective density $\rho_{m}$, radial pressure $p_{r}$, tangential pressure $p_{t}$ and the heat flux components $n_{\alpha}q^{\alpha}$ and $q_{\alpha}q^{\alpha}$. For an anisotropic configurations, the syzygy enforces a nontrivial constraint through the vanishing of ${\bf \it M}$ (as derived in \cite{syzy3, syzy4})
\onecolumngrid
\begin{eqnarray}\label{mastereq}\nonumber
&&{\bf \it M} \equiv -p_{t}^{2}\rho_{m}^{4} - p_{r}^{2}\rho_{m}^{4} + 2p_{r}p_{t}\rho_{m}^{4} - 2p_{t}^{3}\rho_{m}^{3} - 4p_{t}(n_{\alpha}q^{\alpha})^2\rho_{m}^{3} + 4(n_{\alpha}q^{\alpha})^{2}p_{r}\rho_{m}^{3} + 2q_{\alpha}q^{\alpha}p_{t}\rho_{m}^{3} - 2q_{\alpha}q^{\alpha}p_{r}\rho_{m}^{3} - 2p_{r}^{3}\rho_{m}^{3} \\&&\nonumber
+ 2p_{r}^{2}p_{t}\rho_{e}^{3} + 2p_{r}p_{t}^{2}\rho_{m}^{3} -(q_{\alpha}q^{\alpha})^{2}\rho_{m}^{2} - 2p_{r}^{3}p_{t}\rho_{m}^{2} + 6(n_{\alpha}q^{\alpha})^{2}p_{r}^{2}\rho_{m}^{2} - p_{r}^{4}\rho_{m}^{2} + 8q_{\alpha}q^{\alpha}p_{t}^{2}\rho_{m}^{2} + 2q_{\alpha}q^{\alpha}p_{r}^{2}\rho_{m}^{2} - 2p_{r}p_{t}^{3}\rho_{m}^{2} \\&&\nonumber
- 6p_{t}^{2}(n_{\alpha}q^{\alpha})^{2}\rho_{m}^{2} - p_{t}^{4}\rho_{m}^{2} - 10q_{\alpha}q^{\alpha}p_{r}p_{t}\rho_{m}^{2} + 6p_{r}^{2}p_{t}^{2}\rho_{m}^{2} + 8(q_{\alpha}q^{\alpha})^{2}p_{r}\rho_{m} - 30p_{t}^{2}(n_{\alpha}q^{\alpha})^{2}p_{r}\rho_{m} + 2q_{\alpha}q^{\alpha}p_{t}^{3}\rho_{m} \\&&\nonumber
+ 6p_{t}^{3}(n_{\alpha}q^{\alpha})^{2}\rho_{m} + 8q_{\alpha}q^{\alpha}p_{r}^{3}\rho_{m} + 2p_{r}^{3}p_{t}^{2}\rho_{m} + 30p_{r}^{2}(n_{\alpha}q^{\alpha})^{2}p_{t}\rho_{m} + 2p_{t}^{3}p_{r}^{2}\rho_{m} + 18q_{\alpha}q^{\alpha}p_{t}(n_{\alpha}q^{\alpha})^{2}\rho_{m} - 6p_{r}^{3}(n_{\alpha}q^{\alpha})^{2}\rho_{m} \\&&\nonumber
- 2p_{r}p_{t}^{4}\rho_{m} + 10q_{\alpha}q^{\alpha}p_{t}^{2}p_{r}\rho_{m} - 10(q_{\alpha}q^{\alpha})^{2}p_{t}\rho_{m} - 2p_{t}p_{r}^{4}\rho_{m} - 18q_{\alpha}q^{\alpha}(n_{\alpha}q^{\alpha})^{2}p_{r}\rho_{m} - 20p_{r}^{2}q_{\alpha}q^{\alpha}p_{t}\rho_{m} + 2p_{r}^{2}p_{t}^{2}q_{\alpha}q^{\alpha} \\&&\nonumber
- 18q_{\alpha}q^{\alpha}p_{t}^{2}(n_{\alpha}q^{\alpha})^{2} + 10p_{r}^{3}(n_{\alpha}q^{\alpha})^{2}p_{t} - 10p_{t}^{3}(n_{\alpha}q^{\alpha})^{2}p_{r} - p_{t}^{4}p_{r}^{2} - (q_{\alpha}q^{\alpha})^{2}p_{t}^{2} - 4(n_{\alpha}q^{\alpha})^{2}p_{r}^{4} - 8p_{r}^{3}q_{\alpha}q^{\alpha}p_{t} \\&&\nonumber
+ 4q_{\alpha}q^{\alpha}p_{r}^{4} - 8p_{r}(q_{\alpha}q^{\alpha})^{2}p_{t} + 4(q_{\alpha}q^{\alpha})^{3} + 2p_{t}^{3}p_{r}q_{\alpha}q^{\alpha} + 2p_{t}^{3}p_{r}^{3} + 54(n_{\alpha}q^{\alpha})^{2}p_{r}q_{\alpha}q^{\alpha}p_{t} + 4p_{t}^{4}(n_{\alpha}q^{\alpha})^{2} \\&&
- 54(n_{\alpha}q^{\alpha})^{4}p_{r}p_{t} + 27p_{t}^{2}(n_{\alpha}q^{\alpha})^{4} + 27(n_{\alpha}q^{\alpha})^{4}p_{r}^{2} - 36q_{\alpha}q^{\alpha}(n_{\alpha}q^{\alpha})^{2}p_{r}^{2} - p_{r}^{4}p_{t}^{2} + 8(q_{\alpha}q^{\alpha})^{2}p_{r}^{2} = 0.
\end{eqnarray}
\twocolumngrid

The first observation following from Eq. (\ref{mastereq}) is that the the energy-momentum tensor components can not evolve arbitrarily. Their spacetime dependence is constrained to satisfy the polynomial relation encoded in Eq. (\ref{mastereq}) at all values of the coordinates $r$ and $t$. Solving this equation in its full generality is highly nontrivial. In a very recent work, it has been derived that in the context of GR, any singular solution compatible with invariant degeneracy requires the effective matter distribution to always approach isotropy, with anisotropic stresses and energy flux dynamically suppressed \cite{scplb}. As a result, singularity formation can be realized only along a limited class of algebraically admissible matter configurations. However, in modified theories of gravity such algebraic restrictions are richer in structure and the polynomial syzygies are expected to impose analogous algebraic restrictions on the effective matter sector. It is quite possible that curvature-induced and scalar-induced anisotropies enhance or effectively screen the constraints enforced by invariant degeneracy. This raises the possibility that syzygies act not merely as constraints on matter, but as geometric selection rules that can discriminate among admissible modified gravity theories, particularly in the context of a gravitational collapse.  \\

We are interested in a system of collapsing distribution which obeys the standard energy conditions \cite{energy}. The null energy condition requires that for any null vector $k^{\mu}$, $T_{\mu\nu}k^{\mu}k^{\nu} \geq 0$. The weak energy condition states that for any non-spacelike vector $w^{\alpha}$, $T_{\alpha\beta}w^{\alpha}w^{\beta} \geq 0$ and ensures a non-negative energy density for all observers. A dominant energy condition further demands that $-T_{\alpha\beta} w^{\beta}$ be timelike or null for every timelike $w^{\alpha}$ and ensures causal propagation of energy flux. Finally, the strong energy condition is expressed as $2 T_{\alpha\beta} w^{\alpha} w^{\beta} + T \geq 0$, where $w^{\alpha}$ is a unit timelike vector and $T$ is the trace of energy-momentum tensor. We define two different effective equation of state parameters corresponding to radial and tangential pressure and write $p_r = w_1 \rho_m$, $p_t = w_2 \rho_m$. This simplifies Eq. (\ref{mastereq}) into 
\onecolumngrid
\begin{eqnarray}\nonumber
&& \rho_{m}^{6} \Big[w_{2}^2 - w_{1}^2 + 2w_{1}w_{2} - 2w_{2}^3 + 2w_{1}^{2}w_{2} + 2w_{2}^{2}w_{1} - 2w_{1}^{3}w_{2} - w_{1}^{4} - 2w_{1}w_{2}^{3} - w_{2}^{4} + 6w_{1}^{2}w_{2}^{2} + 2w_{1}^{3}w_{2}^{2} + 2w_{2}^{3}w_{1}^{2} - 2w_{1}w_{2}^{4} \\&&\nonumber
- 2w_{2}w_{1}^{4} - w_{2}^{4}w_{1}^{2} + 2w_{1}^{3}w_{2}^{3} - w_{1}^{4}w_{2}^{2} - 2w_{1}^{3} \Big] + \rho_{m}^{4} \Big[-4\Big(n_{\alpha}q^{\alpha}\Big)^{2}w_{2} + 4\Big(n_{\alpha}q^{\alpha}\Big)^{2}w_{1} + 2\Big(q_{\alpha}q^{\alpha}\Big)w_{2} - 2\Big(q_{\alpha}q^{\alpha}\Big)w_{1} \\&&\nonumber
+ 6\Big(n_{\alpha}q^{\alpha}\Big)^{2}w_{1}^{2} + 8\Big(q_{\alpha}q^{\alpha}\Big)w_{2}^{2} + 4\Big(q_{\alpha}q^{\alpha}\Big)w_{1}^{2} - 6\Big(n_{\alpha}q^{\alpha}\Big)^{2}w_{2}^{2} - 10\Big(q_{\alpha}q^{\alpha}\Big)w_{1}w_{2} - 30\Big(n_{\alpha}q^{\alpha}\Big)^{2}w_{1}w_{2}^{2} + 2\Big(q_{\alpha}q^{\alpha}\Big)w_{2}^{3} \\&&\nonumber
+ 6\Big(n_{\alpha}q^{\alpha}\Big)^{2}w_{2}^{3} + 8\Big(q_{\alpha}q^{\alpha}\Big)w_{1}^{3} + 30\Big(n_{\alpha}q^{\alpha}\Big)^{2}w_{1}^{2}w_{2} - 6\Big(n_{\alpha}q^{\alpha}\Big)^{2}w_{1}^{3} + 10\Big(q_{\alpha}q^{\alpha}\Big)w_{2}^{2}w_{1} - 20\Big(q_{\alpha}q^{\alpha}\Big)w_{1}^{2}w_{2} + 10\Big(n_{\alpha}q^{\alpha}\Big)^{2}\\&&\nonumber
w_{2}w_{1}^{3} - 10\Big(n_{\alpha}q^{\alpha}\Big)^{2}w_{2}^{3}w_{1} - 4\Big(n_{\alpha}q^{\alpha}\Big)^{2}w_{1}^{4} - 8\Big(q_{\alpha}q^{\alpha}\Big)w_{1}^{3}w_{2} + 4\Big(q_{\alpha}q^{\alpha}\Big)w_{1}^{4} + 2\Big(q_{\alpha}q^{\alpha}\Big)w_{2}^{3}w_{1} + 4\Big(n_{\alpha}q^{\alpha}\Big)^{2}w_{2}^{4} \Big] \\&&\nonumber
+ \rho_{m}^{2} \Big[-\Big(q_{\alpha}q^{\alpha}\Big)^{2} + 8\Big(q_{\alpha}q^{\alpha}\Big)^{2}w_{1} + 18 \Big(q_{\alpha}q^{\alpha}\Big)\Big(n_{\alpha}q^{\alpha}\Big)^{2}w_{2} - 10\Big(q_{\alpha}q^{\alpha}\Big)^{2}w_{2} - 18 \Big(q_{\alpha}q^{\alpha}\Big)\Big(n_{\alpha}q^{\alpha}\Big)^{2}w_{2}^{2} - \Big(q_{\alpha}q^{\alpha}\Big)^{2}w_{2}^{2} \\&&\nonumber
- 8\Big(q_{\alpha}q^{\alpha}\Big)^{2}w_{1}w_{2} + 54\Big(q_{\alpha}q^{\alpha}\Big)\Big(n_{\alpha}q^{\alpha}\Big)^{2}w_{1}w_{2} - 54\Big(n_{\alpha}q^{\alpha}\Big)^{4}w_{1}w_{2} + 27\Big(n_{\alpha}q^{\alpha}\Big)^{4}w_{2}^{2} + 27\Big(n_{\alpha}q^{\alpha}\Big)^{4}w_{1}^{2} - 36\Big(q_{\alpha}q^{\alpha}\Big)\\&&\label{master2}
\Big(n_{\alpha}q^{\alpha}\Big)^{2}w_{1}^{2} + 8\Big(q_{\alpha}q^{\alpha}\Big)^{2}w_{1}^{2}\Big] + 4\Big(q_{\alpha}q^{\alpha}\Big)^{3} = 0.
\end{eqnarray}
\twocolumngrid

This equation is cubic in the variable $x = \rho_{m}^{2}$ and can be written in the canonical form $ax^{3} + bx^{2} + cx + d = 0$. If $p$, $q$ and $r$ denote the three roots of this polynomial equation, they satisfy the standard relations
\begin{eqnarray}\label{con1}
&& p + q + r = -\frac{b}{a}, \\&& \label{con2}
pq + qr + rp = \frac{c}{a}, \\&& \label{con3}
pqr = -\frac{d}{a}.
\end{eqnarray}

Comparing the roots of Eq. (\ref{master2}) with Eqs. (\ref{con1})-(\ref{con3}), it follows that one or more admissible solutions of effective density can diverge only if the coefficient of the highest power of $\rho_{m}$ (the term corresponding to the coefficient $a$ in $ax^{3} + bx^{2} + cx + d = 0$) goes to zero. This leads to the condition
\begin{eqnarray}\nonumber
&& \Big[w_{2}^2 - w_{1}^2 + 2w_{1}w_{2} - 2w_{2}^3 + 2w_{1}^{2}w_{2} + 2w_{2}^{2}w_{1} - 2w_{1}^{3}w_{2} \\&&\nonumber
- w_{1}^{4} - 2w_{1}w_{2}^{3} - w_{2}^{4} + 6w_{1}^{2}w_{2}^{2} + 2w_{1}^{3}w_{2}^{2} + 2w_{2}^{3}w_{1}^{2} - \\&&\nonumber
2w_{1}w_{2}^{4} - 2w_{2}w_{1}^{4} - w_{2}^{4}w_{1}^{2} + 2w_{1}^{3}w_{2}^{3} - w_{1}^{4}w_{2}^{2} - 2w_{1}^{3} \Big] = 0 .
\end{eqnarray}
This relation provides a general algebraic constraint between the effective radial and tangential pressure components of a self-gravitating sphere in any modified theory of gravity, if we expect the energy density to diverge at a spacetime singularity. As a simple illustration, setting $w_{2} = 0$ (vanishing tangential pressure) reduces the above expression into $w_{1}^{2}(w_{1}^{2}+2w_{1}+1) = 0$. This admits two distinct possibilities : $w_{1}=0$, corresponding to an isotropic dust configuration, or $w_{1} = -1$, for which the radial pressure exhibits a dark-energy like behavior. An analogous conclusion follows upon setting $w_{1} = 0$ (vanishing radial pressure) and solving for $w_{2}$. \\

It is also reasonable to expect that, towards the final stages of a gravitational collapse, when successive matter shells accumulate near the center, the relation between pressure and energy density may deviate from being strictly linear. We emphasize that even for nonlinear equations of state, the restriction imposed by the polynomial syzygy retains the same qualitative form. To illustrate this, we assume a simple quadratic dependence of the pressures on the matter density, as in $p_{r} = \beta_{1}\rho_{m} + \gamma_{1}\rho_{m}^{2}$ and $p_{t} = \beta_{2}\rho_{m} + \gamma_{2}\rho_{m}^{2}$. Substituting this into Eq. (\ref{mastereq}) and retaining the dominant contributions in a density-driven end state, we find that the highest powers of $\rho_{m}$ appearing in the resulting expression are $\rho_{m}^{12}$, $\rho_{m}^{11}$ and $\rho_{m}^{10}$. The constraint equation therefore reduces to a cubic polynomial in $\rho_{m}^{4}$. In the context of cubic equations, the behavior of the roots is governed by the coefficient of the highest power. In the present case, the coefficient of $\rho_{m}^{12}$ term is $\gamma_{1}^{2}\gamma_{2}^{4} + \gamma_{1}^{4}\gamma_{2}^{2} - 2\gamma_{1}^{3}\gamma_{2}^{3}$ and the sum of the roots of the cubic can be expressed as
\onecolumngrid
\begin{equation}
\left(\rho_{m1}^{4} + \rho_{m2}^{4} + \rho_{m3}^{4}\right)
\simeq
\frac{
2\gamma_{1}\gamma_{2}
\left[
\gamma_{1}\gamma_{2}\left(\gamma_{1}+\gamma_{2}\right)
-
\gamma_{2}^{3}\left(1+\beta_{1}\right)
-
\gamma_{1}^{3}\left(1+\beta_{2}\right)
+
\gamma_{1}\gamma_{2}^{2}\left(3\beta_{1}-2\beta_{2}\right)
-
\gamma_{1}^{2}\gamma_{2}\left(2\beta_{1}-3\beta_{2}\right)
\right]
}{
\gamma_{1}^{2}\gamma_{2}^{2}\left(\gamma_{1}-\gamma_{2}\right)^{2}
}.
\end{equation}
\twocolumngrid

From Eqs. (\ref{con1})-(\ref{con3}), it follows that the same denominator $\gamma_{1}^{2}\gamma_{2}^{2}(\gamma_{1}-\gamma_{2})^{2}$ appears in the remaining algebraic relations governing the roots. Consequently, one or more roots can diverge at a singular configuration if and only if this denominator vanishes, precisely when $\gamma_{1}=\gamma_{2}$. This condition corresponds to isotropy in the highest-order density contribution to the pressure components. Thus the argument extends in a straight-forward manner even in the presence of nonlinear equations of state : divergence of the energy density at a spacetime singularity is compatible only if the leading-order pressure terms evolve towards isotropy. 

\section{$f(R)$ Theory of Gravity}\label{fR}
For a chosen modified theory of gravity, the energy-momentum tensor components carry additional theory-induced anisotropies. We expect these to enhance/nullify the constraints enforced by the polynomial degeneracy. We explore our expectation for a spherically symmetric self-gravitating object, described by the metric tensor 
\begin{equation}\label{genmetric}
ds^2 = - A(t,r) dt^2 + B(t,r) dr^2 + C^2(t,r) d\Omega^2,
\end{equation}
in different modified theories of gravity. For a \emph{metric $f(R)$ gravity}, the standard field equations are obtained by varying the action with respect to the metric tensor 
\begin{equation}
f_R R_{\mu\nu} - \frac{1}{2} f g_{\mu\nu} + (g_{\mu\nu}\Box - \nabla_\mu\nabla_\nu)f_R = 8\pi T_{\mu\nu},
\end{equation}
where $f_R \equiv df/dR$. Dividing the above equation by $f_R$ and rearranging, one can write it in an effective Einstein form
\begin{equation}
G_{\mu\nu} = 8\pi T^{\rm eff}_{\mu\nu},
\end{equation}
where the effective stress-energy tensor is given by
\begin{eqnarray}\nonumber
&& T^{\rm eff}_{\mu\nu} = \frac{1}{f_R}T_{\mu\nu}+\frac{1}{8\pi f_R} \Big[\nabla_\mu\nabla_\nu f_R-g_{\mu\nu}\Box f_R + \frac{1}{2}(f \\&&
- R f_R) g_{\mu\nu} \Big].
\end{eqnarray}
This expression simply portrays the well-known fact that $f(R)$ gravity can be interpreted as GR coupled to an effective fluid composed of matter and curvature degrees of freedom \cite{fR1, fR2, fR3}. For the metric in Eq. (\ref{genmetric}), the effective anisotropy is
\begin{eqnarray}\nonumber\label{anisofR}
&& p_r^{\rm eff} - p_t^{\rm eff} = \frac{p_r - p_t}{f_R} + \frac{1}{8\pi f_R} \Bigg[f_R'' - \left(\frac{B'}{2B}+\frac{C C'}{B}\right)f_R' \\&&
 + \left(\frac{C\dot C}{A}-\frac{\dot B}{2A}\right)\dot f_R \Bigg].
\end{eqnarray}
Eq. \eqref{anisofR} makes it explicit that, even if the physical matter sector is isotropic ($p_r = p_t$), curvature contributions introduce an effective anisotropy. The polynomial degeneracy demands that at singular roots, the trace-free Ricci anisotropy must vanish. In the context of $f(R)$, this requirement translates into
\begin{equation}
p_r^{\rm eff}-p_t^{\rm eff} = 0.
\label{syzygyfR}
\end{equation}

Using the above expression we rewrite Eq. (\ref{anisofR}) as
\begin{equation}
p_r-p_t = -\frac{1}{8\pi} \Bigg[ f_R'' - \left(\frac{B'}{2B}+\frac{C C'}{B}\right)f_R' + \left(\frac{C\dot C}{A}-\frac{\dot B}{2A}\right)\dot f_R \Bigg].
\label{balancefR}
\end{equation}

Eq. \eqref{balancefR} allows us to frame the following selection rule : singularity formation is possible only if matter anisotropy dynamically compensates curvature-induced anisotropy. A generic $f(R)$ collapse tends to violate this balance and naturally disfavors a formation of singularity unless heavily constrained in the lagrangian itself. For a general function $f(R)$, the curvature contribution to the effective anisotropy is proportional to
\begin{align}
\nabla_r \nabla_r f_R - \nabla_\theta \nabla_\theta f_R &= f_{RR} \left( \nabla_r \nabla_r R - \nabla_\theta \nabla_\theta R \right)
\nonumber\\
&\quad+f_{RRR}\left[(\nabla_r R)^2 - (\nabla_\theta R)^2\right].
\label{chainrulefR}
\end{align}
Near the collapsing core, curvature gradients grow rapidly and the second-derivative terms dominate over first-derivatives. In this limit, the leading-order behavior of the effective anisotropy is controlled by the combination
\begin{equation}
\frac{f_{RR}}{f_R}\left(\nabla_r \nabla_r R - \nabla_\theta \nabla_\theta R \right),
\end{equation}
which provides a reliable approximation of the curvature-induced anisotropy close to a singular end-state. As a simple example, we consider the Starobinsky model $f(R) = R + \alpha R^2$ \cite{fR1}, for which $f_R = 1 + 2\alpha R$. Substituting into Eq. \eqref{anisofR}, we find that
\begin{equation}
p_r^{\rm eff}-p_t^{\rm eff} = \frac{p_r-p_t}{1+2\alpha R} + \frac{\alpha}{4\pi(1+2\alpha R)} \left(
\nabla_r\nabla_r R - \nabla_\theta\nabla_\theta R \right).
\end{equation}

Near a collapsing core with high, the second term becomes dominant unless $\alpha$ is sufficiently small or curvature gradients are suppressed. For $\alpha \gg 0$, the curvature contribution enhances effective anisotropy, thereby obstructing the vanishing condition in Eq. \eqref{syzygyfR}. Consequently, singularity formation requires fine-tuned matter anisotropy or a breakdown of the collapse symmetry. This suggests a requirement of the qualitative bound
\begin{equation}
\alpha R \ll 1
\end{equation}
as a necessary (though not sufficient) condition for singularity formation. When $\alpha R \gtrsim 1$, the curvature-induced anisotropy dominates, favoring bounce or a regularized collapse. To assess the behavior of curvature-induced anisotropy near the central collapsing shell, we consider the limit $r \to 0$, assuming that the metric coefficients approximately behave as $A(t,r) \sim A_0(t)$, $B(t,r) \sim B_0(t)$, $C(t,r) \sim r a(t)$. The Ricci divergence comes as
\begin{equation}
R(t,r) \sim r^{-q}, \qquad q > 0.
\end{equation}

The radial derivatives of the areal radius scale as
\begin{equation}
C' \sim a(t), \qquad C'' \sim 0 .
\end{equation}

Deriving the effective contributions in the curvature-driven anisotropy for all $q > 0$, we find that the dominant divergence arises due to a scaling of
\begin{equation}
\nabla_r\nabla_r R - \nabla_\theta\nabla_\theta R \;\sim\; r^{-(3q+2)}, \qquad r\to0 .
\label{scalingnablaR}
\end{equation}

As a consequence, the leading curvature-induced anisotropy entering the syzygy constraint scales as
\begin{equation}\label{scalingnablaRexact}
\frac{1}{f_R}\left(\nabla_r \nabla_r f_R - \nabla_\theta \nabla_\theta f_R\right) \;\sim\; \frac{f_{RR}}{f_R}\, r^{-(3q+2)}.
\end{equation}

Eq. (\ref{scalingnablaRexact}) suggests that even if the fluid inside a collapsing sphere is isotropic, the curvature induced anisotropy in $f(R)$ gravity is usually non-zero unless $f_{RR}$ is very small. The other passage to reach a zero effective anisotropy remains the same; choosing a very special stress-energy tensor for the fluid anisotropy that simply cancels out the curvature anisotropy. Moreover, it seems that a singularity formation in $f(R)$ gravity also depends on the growth rate of the ratio $f_{RR}/f_R$ at large curvature. In models of $f(R)$ collapse for which this ratio grows sufficiently fast, there is an amplification of effective anisotropy and the vanishing condition required by polynomial degeneracies can not be satisfied. 

\section{Scalar-Tensor Theory}
The analysis with metric $f(R)$ gravity shows that higher-derivative curvature terms generate effective anisotropies and obstruct the degeneracy condition favorable for singularity formation. A natural follow-up question then arises : does a similar obstruction persist when the additional degrees of freedom are encoded not in higher curvature corrections, but in a scalar field non-minimally coupled to geometry? Scalar-tensor theories provide a minimal and structurally transparent framework to address this question \cite{wagoner, fujii}. A scalar-tensor theory in Jordan frame is described by the action
\begin{equation}
S = \int d^4x\sqrt{-g} \left[\frac{1}{2}\mathcal{F}(\phi)R - \frac{1}{2}g^{\mu\nu}\nabla_\mu\phi\nabla_\nu\phi - V(\phi)+ \mathcal{L}_{m} \right],
\end{equation}
where $\mathcal{F}(\phi)$ defines a non-minimal coupling function and $V(\phi)$ is the scalar self-interaction potential. Varying the action with respect to the metric yields
\begin{eqnarray}\nonumber
&&\mathcal{F}(\phi) G_{\mu\nu} = T_{\mu\nu} + \nabla_\mu\phi\nabla_\nu\phi - \frac{1}{2}g_{\mu\nu}(\nabla\phi)^2 - g_{\mu\nu}V(\phi) +\\&& 
\nabla_\mu\nabla_\nu \mathcal{F} - g_{\mu\nu}\Box \mathcal{F},
\label{STfield}
\end{eqnarray}
where $T_{\mu\nu}$ is the stress-energy tensor for ordinary matter. Eq. \eqref{STfield} may be re-written in an effective Einstein-like form as in
\begin{eqnarray}
&& G_{\mu\nu} = 8\pi T^{\rm eff}_{\mu\nu},\\&& \nonumber
T^{\rm eff}_{\mu\nu} = \frac{1}{\mathcal{F}}T_{\mu\nu} + \frac{1}{\mathcal{F}}\left[\nabla_\mu\phi\nabla_\nu\phi - \frac{1}{2}g_{\mu\nu}(\nabla\phi)^2 - g_{\mu\nu}V
\right]\\&&
+ \frac{1}{\mathcal{F}} \left(\nabla_\mu\nabla_\nu \mathcal{F} - g_{\mu\nu}\Box \mathcal{F} \right).
\end{eqnarray}

For a general non-static spherically symmetric spacetime as in Eq. (\ref{genmetric}), the effective anisotropy can be derived from the field equations as
\begin{align}
p_r^{\rm eff}-p_t^{\rm eff}
&=
\frac{p_r-p_t}{\mathcal{F}}
+
\frac{\phi'^2}{B\,\mathcal{F}}
+
\frac{1}{\mathcal{F}}
\left(
\nabla_r\nabla_r\mathcal{F}
-
\nabla_\theta\nabla_\theta\mathcal{F}
\right).
\label{anisost2}
\end{align}

From Eq. (\ref{anisost2}) one can note once again that even if the matter sector is perfectly isotropic ($p_r = p_t$), spatial gradients of the scalar field and second derivatives of the non-minimal coupling function in general induce an effective anisotropy. An explicit form of the coupling-induced anisotropy can be derived as
\begin{equation}
\nabla_r\nabla_r\mathcal{F} - \nabla_\theta\nabla_\theta\mathcal{F} = \mathcal{F}_{,\phi}\left(\nabla_r\nabla_r\phi -
\nabla_\theta\nabla_\theta\phi \right) + \frac{\mathcal{F}_{,\phi\phi}}{B}\phi'^2.
\end{equation}

The polynomial degeneracies of Ricci invariants require the vanishing of the effective trace-free anisotropy at singular roots, i.e., $p_r^{\rm eff} - p_t^{\rm eff} = 0$. Using Eq. \eqref{anisost2}, this yields a constraint for the generalized scalar-tensor theory as
\begin{eqnarray}\nonumber
&& p_r-p_t = -\mathcal{F}_{,\phi}\,\phi'' + \mathcal{F}_{,\phi} \left(\frac{B'}{2B}+\frac{C C'}{B}\right)\phi' \\&&
- \mathcal{F}_{,\phi}\left(\frac{C\dot C}{A}-\frac{\dot B}{2A}\right)\dot\phi - \frac{1+\mathcal{F}_{,\phi\phi}}{B}\,\phi'^2 .
\label{STconstraint}
\end{eqnarray}

Eq. (\ref{STconstraint}) admits three crucial implications for the admissibility of singular end states in scalar-tensor gravity:
\begin{enumerate}
\item In the absence of matter-induced anisotropy ($p_r = p_t$), the syzygy condition enforces a balance between the second-derivative contributions from the non-minimal coupling $\left(\nabla_r\nabla_r\mathcal{F} - \nabla_\theta\nabla_\theta\mathcal{F} \right)$ and the spatial gradient of the scalar field to effectively drive the configuration towards a total isotropy.

\item One may reach a singular root with a spatially homogeneous scalar field and the condition $\mathcal{F}_{,\phi} \to 0$, driving the theory into a minimally coupled theory of scalar field.

\item Inhomogeneous scalar fields for which neither $\phi'$ nor $\phi''$ are dynamically suppressed, the syzygy constraint is generically violated near singular configurations unless finely tuned with respect to matter anisotropy or the functional form of $\mathcal{F}(\phi)$.
\end{enumerate}

Taken together, these results indicate that scalar-tensor theories ca only permit singular end states within a constrained subset of field configurations, much like an $f(R)$ theory. Crucially, this restriction arises from algebraic degeneracies of curvature invariants and their coupling to the scalar sector, rather than the specific choice of a scalar potential $V(\phi)$ or the evolution of the field(s). 

\section{Brans-Dicke Theory}
We now look into the special case of Brans-Dicke (BD) theory \cite{BD}, where the non-minimal coupling and kinetic structure are fixed by symmetry. Brans-Dicke theory occupies a central position within the broader class of scalar-tensor theories, as it provides the simplest and most historically well-motivated framework in which gravity is mediated not only by the spacetime metric but also by a dynamical scalar field. Originally proposed to incorporate Mach's principle into gravitation, the theory introduces a single dimensionless coupling parameter that controls deviations from GR while preserving second-order field equations in the Jordan frame. As such, BD gravity serves as a natural benchmark for testing both conceptual and observational consequences of scalar-tensor modifications, with GR recovered in an appropriate limit \cite{BD1}. Moreover, many generalized scalar-tensor and modified gravity models reduce to BD theory under specific choices of coupling functions or field redefinitions \cite{BD2}, making it an essential reference to understand the structure, consistency and phenomenology of more generic scalar-tensor extensions. The Brans-Dicke action is written as
\begin{equation}
S = \frac{1}{16\pi}\int d^4x\sqrt{-g} \left[\phi R - \frac{\omega}{\phi}g^{\mu\nu}\nabla_\mu\phi\nabla_\nu\phi \right],
\end{equation}
where $\phi$ is the BD scalar field and $\omega$ is the dimensionless BD parameter. A variation with respect to the metric yields
\begin{eqnarray}\nonumber
&& \phi G_{\mu\nu} = 8\pi T_{\mu\nu} + \frac{\omega}{\phi} \left\lbrace\nabla_\mu\phi\nabla_\nu\phi - \frac{1}{2}g_{\mu\nu}(\nabla\phi)^2 \right\rbrace \\&& + \nabla_\mu\nabla_\nu\phi - g_{\mu\nu}\Box\phi.
\end{eqnarray}
As in the previous sections, this equation can be written in Einstein-like form $G_{\mu\nu} = 8\pi T^{\rm eff}_{\mu\nu}$, with
\begin{eqnarray}\nonumber
&& T^{\rm eff}_{\mu\nu} = \frac{1}{\phi}T_{\mu\nu} + \frac{\omega}{8\pi\phi^2} \left(\nabla_\mu\phi\nabla_\nu\phi - \frac{1}{2}g_{\mu\nu}(\nabla\phi)^2 \right)\\&& 
+ \frac{1}{8\pi\phi} \left(\nabla_\mu\nabla_\nu\phi - g_{\mu\nu}\Box\phi\right).
\end{eqnarray}

For a general non-static spherically symmetric metric as in Eq. (\ref{genmetric}), the effective anisotropy can be derived as
\begin{eqnarray}\nonumber
&& p_r^{\rm eff}-p_t^{\rm eff} = \frac{p_r-p_t}{\phi} + \frac{\omega}{8\pi\phi^2}\frac{\phi'^2}{B} + \frac{1}{8\pi\phi} \big(\nabla_r\nabla_r\phi \\&&
- \nabla_\theta\nabla_\theta\phi \big), \\&&\label{BDaniso2}\nonumber
\nabla_r\nabla_r\phi - \nabla_\theta\nabla_\theta\phi = \phi'' - \left(\frac{B'}{2B} + \frac{CC'}{B} \right)\phi' \\&&
+\left(\frac{C\dot C}{A}-\frac{\dot C}{2A} \right)\dot\phi .
\end{eqnarray}

This term is purely geometric and independent of the BD parameter $\omega$. The polynomial degeneracies of Ricci invariants impose the vanishing of the effective trace-free Ricci anisotropy at singular roots,
\begin{equation}
p_r^{\rm eff} - p_t^{\rm eff}=0 .
\label{BDsyzygy}
\end{equation}

Substituting Eq. \eqref{BDaniso2} into Eq. \eqref{BDsyzygy} we find the algebraic constraint
\begin{equation}
p_r-p_t = -\frac{\omega}{8\pi\phi}\frac{\phi'^2}{B}-\frac{1}{8\pi}\left(\nabla_r\nabla_r\phi-\nabla_\theta\nabla_\theta\phi\right).
\label{BDconstraint}
\end{equation}

A few important consequences follow directly. \\

\textbf{(i) Matter isotropy :} \\
If the matter sector is isotropic ($p_r=p_t$), Eq.~\eqref{BDconstraint} implies
\begin{equation}
\frac{\omega}{\phi}\frac{\phi'^2}{B} + \left(\nabla_r\nabla_r\phi - \nabla_\theta\nabla_\theta\phi \right) = 0.
\end{equation}
This condition implies that for generic collapsing geometry to end up in isotropic singularity, $\phi = \phi(t)$ is prefered. \\

\textbf{(ii) Restriction on $\omega$ :} \\
For inhomogeneous scalar configurations ($\phi'\neq 0$), the first term in Eq. \eqref{BDconstraint} is strictly negative for $\omega > 0$. Hence, for positive $\omega$, scalar gradients inevitably generate effective anisotropy that obstructs the syzygy condition. Singular end states are therefore incompatible with generic inhomogeneous BD scalars when $\omega > 0$. On the contrary, $\omega < 0$ can allow partial cancellation between scalar-gradient and second-derivative terms, but at the cost of introducing ghost-like kinetic behavior. Therefore, from the view-point of polynomial degeneracies alone, singularity formation in Brans-Dicke theory favors either a homogeneous scalar field or an effective suppression of the kinetic sector, $\omega \to 0$. \\

\textbf{(iii) GR limit :} \\
In the limit $\omega \to \infty$ with $\phi \to \text{const}$, BD theory reduces to GR. This limit automatically satisfies the syzygy constraint and restores the standard singularity structure of GR. However, any finite $\omega$ admits a restricted subset of collapsing solutions compatible with invariant degeneracy.

\section{Generalized Brans-Dicke Theory}
We have seen that scalar gradients and non-minimal couplings generically act as sources of effective anisotropy, severely restricting the admissibility of singular end states unless the scalar sector becomes homogeneous dynamically. In the standard BD theory, this restriction is particularly stringent : a constant coupling parameter fixes the relative weight of scalar-gradient contributions, leaving little room for compensating anisotropy during collapse. This motivates the consideration of generalized BD theories, where the coupling parameter $\omega$ is promoted to a scalar-dependent function \cite{genBD}. Such a generalization introduces a new dynamical freedom through which scalar-induced anisotropy may be amplified or suppressed. From the perspective of invariant degeneracies, this allows one to investigate whether the running of $\omega(\phi)$ can evade the algebraic obstructions imposed by syzygies. The action we consider is
\begin{equation}
S=\frac{1}{16\pi}\int d^4x\sqrt{-g}\left[\phi R-\frac{\omega(\phi)}{\phi}(\nabla\phi)^2\right].
\end{equation}

Variation with respect to the metric yields
\begin{eqnarray}\nonumber
&& \phi G_{\mu\nu} = 8\pi T_{\mu\nu} + \frac{\omega(\phi)}{\phi}\left(\nabla_\mu\phi\nabla_\nu\phi - \frac{1}{2}g_{\mu\nu}(\nabla\phi)^2 \right) \\&&
+\nabla_\mu\nabla_\nu\phi - g_{\mu\nu}\Box\phi.
\label{GBDmetric}
\end{eqnarray}
As before, we define the effective stress-energy tensor via $G_{\mu\nu} = 8\pi T^{\rm eff}_{\mu\nu}$ and derive the effective anisotropy for the non-static spherically symmetric metric as in Eq. (\ref{genmetric})
\begin{eqnarray}\nonumber
&& p_r^{\rm eff}-p_t^{\rm eff} = \frac{p_r-p_t}{\phi} + \frac{\omega(\phi)}{8\pi\phi^2}\frac{\phi'^2}{B} + \frac{1}{8\pi\phi} \big(\nabla_r\nabla_r\phi \\&&
- \nabla_\theta\nabla_\theta\phi \big).
\label{GBDaniso1}
\end{eqnarray}

The generalized Brans-Dicke scalar evolution equation is
\begin{equation}
\Box\phi = \frac{8\pi T}{3+2\omega(\phi)} - \frac{d\omega / d\phi}{3+2\omega(\phi)}(\nabla\phi)^2 ,
\label{GBDscalar}
\end{equation}
where $T$ denotes the trace of the matter energy-momentum tensor. The d'Alembertian takes the elaborate form
\begin{eqnarray}\nonumber
&& \Box\phi = -\frac{1}{A} \Big(\ddot\phi - \frac{\dot A}{2A}\dot\phi + \frac{A'}{2B}\phi' \Big) + \frac{1}{B} \Big(\phi'' - \frac{B'}{2B}\phi' \\&&
+ \frac{\dot B}{2A}\dot\phi \Big) + \frac{2}{C^2}\Big(\frac{C C'}{B}\phi' - \frac{C\dot C}{A}\dot\phi \Big).
\label{boxphi_expanded}
\end{eqnarray}

Deriving the scalar kinetic invariant $(\nabla\phi)^2$ and substituting Eq. \eqref{boxphi_expanded} in Eq. \eqref{GBDscalar}, we isolate the second radial derivative $\phi''$ to write
\begin{equation}\label{phidprime}
\frac{1}{B}\phi'' = \frac{8\pi T}{3+2\omega(\phi)} - \frac{d\omega / d\phi}{3+2\omega(\phi)} \left( -\frac{\dot\phi^2}{A}
+ \frac{\phi'^2}{B} \right) + \mathcal{S}(t,r),
\end{equation}
where $\mathcal{S}(t,r)$ is a collection of terms involving first derivatives of $\phi$. To derive an explicit expression of the effective anisotropy, we write
\begin{eqnarray}\nonumber
&& \nabla_r\nabla_r\phi - \nabla_\theta\nabla_\theta\phi = \phi'' - \left(\frac{B'}{2B} + \frac{C C'}{B} \right)\phi'
\\&&
+ \left(\frac{C\dot C}{A} - \frac{\dot B}{2A} \right)\dot\phi.
\label{2nddiffphi}
\end{eqnarray}

Now using Eq. \eqref{phidprime} and \eqref{2nddiffphi} in Eq. \eqref{GBDaniso1} we write
\begin{eqnarray}\nonumber
&&p_r^{\rm eff}-p_t^{\rm eff} = \frac{p_r-p_t}{\phi} + \frac{\omega(\phi)}{8\pi\phi^2}\frac{\phi'^2}{B} + \frac{1}{8\pi\phi}\Big[\frac{B}{3+2\omega(\phi)}\\&&
\left(8\pi T - \frac{d\omega}{d\phi}\frac{\phi'^2}{B}\right)\Big] + \frac{1}{8\pi\phi} B \mathcal{S}(t,r).
\end{eqnarray}

If we assume that near a spacelike singularity, the spatial gradients of the geometric scalar field dominate, then $\frac{\phi'^2}{B} \gg T$ and $\frac{\phi'^2}{B} \gg \mathcal{S}(t,r)$. Retaining only the leading order divergence, the effective anisotropy reduces to
\begin{equation}
p_r^{\rm eff}-p_t^{\rm eff} \simeq \frac{p_r-p_t}{\phi} + \frac{\phi'^2}{B} \left[\frac{\omega(\phi)}{8\pi\phi^2} - \frac{d\omega / d\phi}{8\pi\phi\,[3+2\omega(\phi)]}\right].
\end{equation}

Polynomial degeneracies of Ricci invariants require the vanishing of the effective trace-free Ricci anisotropy at singular roots, i.e., $p_r^{\rm eff} - p_t^{\rm eff} = 0$. If ordinary matter is isotropic, this requirement becomes
\begin{equation}\label{omegacond}
\frac{\omega(\phi)}{\phi^2} - \frac{\omega'(\phi)}{\phi [3+2\omega(\phi)]} = 0.
\end{equation}

Eq. \eqref{omegacond} represents a purely geometric restriction on the functional form of the Brans-Dicke coupling, arising from invariant degeneracy rather than phenomenological considerations. It admits power-law solutions of the form
\begin{equation}
\omega(\phi) \sim \phi^n, \qquad n > 1,
\end{equation}
for which the scalar-gradient contributions to effective anisotropy can cancel out dynamically. This result has a few important implications

\begin{itemize}
\item A standard Brans-Dicke theory with constant $\omega$ generically obstructs singularity formation due to scalar-induced anisotropy.
\item A running coupling $\omega(\phi)$ provides a natural mechanism for anisotropy screening without enforcing scalar homogeneity near singularity.
\item The GR limit is recovered asymptotically when $\omega(\phi) \to \infty$, consistent with both weak-field phenomenology and strong-field syzygy constraints.
\end{itemize}

Thus, generalized Brans-Dicke theory emerges as the minimal scalar-tensor extension capable of evading the invariant-induced obstruction to a singular collapse.

\section{Chameleon Brans-Dicke Theory}
The generalized BD framework demonstrates that allowing the coupling parameter $\omega(\phi)$ to evolve dynamically introduces a potential mechanism for regulating scalar-induced anisotropy during gravitational collapse. Perhaps a better motivated extension can also be provided by a chameleon Brans-Dicke theory, where the scalar field couples directly to the matter sector through a field-dependent coupling. This interaction renders the scalar dynamics explicitly environment-dependent, enabling the scalar field to respond to local density variations \cite{khoury, hinter, scdutta}. From the standpoint of invariant degeneracies, such matter couplings introduce an additional geometric channel through which scalar-gradient anisotropy may be screened or amplified. A chameleon Brans-Dicke theory is described by the action \cite{scdutta}
\begin{eqnarray}\label{1}\nonumber
&& S = \frac{1}{16\pi}\int d^4x \sqrt{-g}\Big[ \phi R - \frac{\omega_{bd}}{\phi}g^{\mu\nu}\nabla_{\mu}\phi \nabla_{\nu}\phi - V(\phi) \\&& 
+ 16\pi f(\phi)L_m\Big].
\end{eqnarray}
Here $f(\phi)$ represents the nonminimal coupling between the scalar field and matter. Varying the action with respect to the metric yields
\begin{eqnarray}\nonumber
&&\phi G_{\mu\nu} = 8\pi f(\phi)\,T_{\mu\nu} + \frac{\omega_{bd}}{\phi}\left(\nabla_\mu \phi \nabla_\nu\phi - \frac{1}{2}g_{\mu\nu}(\nabla\phi)^2 \right) \\&&
+ \nabla_\mu \nabla_\nu\phi - g_{\mu\nu}\Box\phi -\frac{1}{2}g_{\mu\nu}V(\phi).
\label{CBDmetric}
\end{eqnarray}

Dividing the field equations by $\phi$, the gravitational dynamics may be written in an effective Einstein form, $G_{\mu\nu} = 8\pi T^{\rm eff}_{\mu\nu}$ where the effective stress-energy tensor is defined as
\begin{equation}
T^{\rm eff}_{\mu\nu} = \frac{f(\phi)}{\phi}T_{\mu\nu} + T^{(\phi)}_{\mu\nu},
\end{equation}
with $T^{(\phi)}_{\mu\nu}$ collecting all contributions arising from the scalar field. Variation of the action with respect to the scalar field $\phi$ yields the scalar equation of motion,
\begin{eqnarray}\nonumber
&& \Box\phi = \frac{8\pi f(\phi)}{3+2\omega_{bd}} T - \frac{2(df/d\phi)}{3+2\omega_{bd}} L_m + \frac{1}{3+2\omega_{bd}}
\Big\lbrace \phi \frac{dV}{d\phi} \\&&
- 2V(\phi)\Big\rbrace.
\label{CBDscalar}
\end{eqnarray}
Near the end stages of gravitational collapse, the dynamics is expected to be density dominated. For ordinary matter, we approximate $T \simeq -\rho_m$, retain only the leading density-dependent contribution and simplify the scalar field equation into
\begin{equation}
\Box\phi \sim \frac{8\pi\rho_m}{3+2\omega_{bd}} \left[- f(\phi)+\phi \frac{df}{d\phi}\right].
\end{equation}

Following similar steps shown in the last section, we derive the leading order scalar-induced anisotropy factor as
\begin{equation}
\Delta_\phi^{(\rm dom)} = \frac{\phi'^2}{B} \left[\frac{\omega_{bd}}{8\pi\phi^2} - \frac{(df/d\phi)}{8\pi\phi(3+2\omega_{bd})} \right].
\label{CBDanisograd}
\end{equation}

Polynomial degeneracies of Ricci invariants require that $p_r^{\rm eff} - p_t^{\rm eff} = 0$. For isotropic matter distributions, this condition reduces to the requirement of $\Delta_\phi^{(\rm dom)} = 0$. Using Eq. (\ref{CBDanisograd}), this yields the algebraic condition that the chameleon coupling function must satisfy
\begin{equation}
\frac{df}{d\phi} = \frac{3+2\omega_{bd}}{\phi} \omega_{bd}.
\end{equation}

This is a purely geometric restriction on the scalar-matter coupling arising from invariant degeneracy. Singularity formation in chameleon Brans-Dicke gravity is therefore admissible only for a restricted functional class of coupling functions $f(\phi)$. Generic choices of $f(\phi)$ tend to generate unavoidable scalar-induced anisotropy, violating the syzygy constraint and dynamically obstructing singular end states.

\section{Syzygies and the Raychaudhuri equation}
The Raychaudhuri equation provides a purely geometric description of the kinematic evolution of timelike congruences and plays a central role in the analysis of gravitational focusing and a formation of singularity \cite{rc}. We have seen that the polynomial degeneracies of curvature invariants impose algebraic constraints on the trace-free part of the Ricci tensor, which can be imagined as a source of shear in an evolving congruence. Therefore, syzygies can restrict the admissible shear dynamics, independent of the field equations of a theory. We use this to establish a natural bridge between invariant degeneracies of Riemannian geometry and the Raychaudhuri framework : singularity formation is permitted only when the algebraic structure of curvature invariants allows sustained focusing without obstruction from shear-induced defocusing. Let $u^\mu$ be a unit time-like vector field tangent to a congruence of worldlines, $u^\mu u_\mu=-1$. The covariant derivative of $u^\mu$ can be decomposed into
\begin{equation}
\nabla_\nu u_\mu = \frac{1}{3}\theta\,h_{\mu\nu} + \sigma_{\mu\nu} + \omega_{\mu\nu} - a_\mu u_\nu,
\label{uDecomp}
\end{equation}
where $\theta \equiv \nabla_\mu u^\mu$ is the expansion scalar, $\sigma_{\mu\nu}$ is the shear tensor, $\omega_{\mu\nu}$ is the vorticity tensor, $a_\mu \equiv u^\nu\nabla_\nu u_\mu$ is the four-acceleration and $h_{\mu\nu} \equiv g_{\mu\nu} + u_\mu u_\nu$ projects orthogonally to $u^\mu$. The shear tensor is symmetric, trace-free and spatial, i.e.,
\begin{equation}
\sigma_{\mu\nu}=\sigma_{\nu\mu}, \qquad
\sigma^\mu{}_\mu=0, \qquad
\sigma_{\mu\nu}u^\nu=0 .
\end{equation}

The evolution equation for $\sigma_{\mu\nu}$ is derived from the Ricci identity applied to $u^\mu$,
\begin{equation}
\nabla_{[\alpha}\nabla_{\beta]}u_\mu = R_{\alpha\beta\mu\nu}u^\nu,
\label{RicciIdentity}
\end{equation}
followed by orthogonal projection to $u^\mu$ and extraction of the symmetric trace-free part. After a standard but lengthy calculation, one can derive the shear propagation equation as
\begin{eqnarray}\nonumber
&&\dot\sigma_{\mu\nu} \equiv u^\lambda\nabla_\lambda\sigma_{\mu\nu}\\&&\nonumber
=-\frac{2}{3}\theta \sigma_{\mu\nu}-\sigma_{\mu\lambda}\sigma^\lambda{}_\nu - \omega_{\mu\lambda}\omega^\lambda{}_\nu +
\frac{1}{3}h_{\mu\nu}\big(\sigma_{\alpha\beta}\sigma^{\alpha\beta} - \omega_{\alpha\beta}\\&&\nonumber
\omega^{\alpha\beta} \big)- E_{\mu\nu} - \frac{1}{2} \left(h_\mu{}^\alpha h_\nu{}^\beta - \frac{1}{3}h_{\mu\nu}h^{\alpha\beta} \right) R_{\alpha\beta} + \nabla_{\langle\mu}a_{\nu\rangle}, \\&&\nonumber
E_{\mu\nu} \equiv C_{\mu\alpha\nu\beta}u^\alpha u^\beta.
\end{eqnarray}
Angular brackets denote the projected, symmetric, trace-free part. Using the Ricci decomposition
\begin{equation}
R_{\mu\nu} = \frac{1}{4}R\,g_{\mu\nu} + S_{\mu\nu} ~,~ S^\mu{}_\mu=0,
\end{equation}
the projected trace-free Ricci term reduces exactly to
\begin{equation}
\left(h_\mu{}^\alpha h_\nu{}^\beta - \frac{1}{3}h_{\mu\nu}h^{\alpha\beta} \right)R_{\alpha\beta} = S_{\mu\nu}.
\end{equation}

Assuming geodesic flow ($a_\mu=0$) for hypersurface-orthogonal congruences ($\omega_{\mu\nu} = 0$) we can write the shear propagation equation as
\begin{equation}
\dot\sigma_{\mu\nu} + \frac{2}{3}\theta \sigma_{\mu\nu} = - E_{\mu\nu} - \frac{1}{2}S_{\mu\nu} - \sigma_{\mu\lambda}\sigma^\lambda{}_\nu + \frac{1}{3}h_{\mu\nu}\sigma_{\alpha\beta}\sigma^{\alpha\beta}.
\label{ShearFinal}
\end{equation}

We recall that the polynomial syzygies among Ricci invariants impose algebraic relations between scalar contractions of $S_{\mu\nu}$. In particular, the quadratic invariant $r_1 \equiv S_{\mu\nu}S^{\mu\nu}$ measures the invariant magnitude of the trace-free Ricci tensor.From Eq. \eqref{ShearFinal} we note that the trace-free Ricci tensor $S_{\mu\nu}$ acts as a direct algebraic source for shear, much like the Weyl curvature tensor. As a consequence, any invariant constraint that enforces $S_{\mu\nu} \to 0$ eliminates a fundamental source of shear in the evolution of a given congruence. The shear evolution equation then reduces to a simple balance between the kinematic damping induced by the expansion scalar and the remaining terms. In the collapse scenarios of interest, the expansion scalar satisfies $\theta \to -\infty$ as the congruence approaches a focusing singularity. Consequently, the term $(2/3)\theta \sigma_{\mu\nu}$ acts as a strong dissipative contribution, driving any residual shear towards decay. At the same time, spherical symmetry and proximity to syzygy-compatible configurations constrain the electric part of the Weyl tensor $E_{\mu\nu}$ to be sub-leading compared to Ricci-sourced effects, while the quadratic shear terms are unable to sustain growth in the absence of a source. Taken together, these features imply that, although $S_{\mu\nu} \to 0$ does not instantaneously enforce $\sigma_{\mu\nu} = 0$ as an algebraic identity, it removes the only invariant mechanism capable of maintaining shear during collapse. The subsequent evolution is therefore governed by dissipative dynamics, leading to an invariant suppression of the shear scalar. This result highlights that polynomial degeneracies of curvature invariants dynamically restrict the kinematical distortion of timelike congruences by eliminating invariant sources of anisotropic deformation.  \\

The Raychaudhuri equation governing a hypersurface-orthogonal timelike congruence with tangent vector $u^\mu$ is
\begin{equation}
\frac{d\theta}{d\tau} = - \frac{1}{3}\theta^2 - \sigma^2 - R_{\mu\nu}u^\mu u^\nu.
\label{Raychaudhuri}
\end{equation}
At a syzygy-compatible singular root, polynomial degeneracies enforce the invariant suppression of the trace-free Ricci tensor, $S_{\mu\nu} \to 0$, and as already discussed, this removes the Ricci-sourced production of shear along the collapsing congruence. As a result, the shear term in Eq. \eqref{Raychaudhuri} becomes sub-leading near the singularity. Moreover, once $S_{\mu\nu} = 0$, the Ricci tensor becomes purely isotropic, i.e., $R_{\mu\nu}u^\mu u^\nu = \frac{1}{4}R$, independent of the choice of congruence. Substituting these into Eq. \eqref{Raychaudhuri}, we write the simplified equation near a syzygy-compatible singular configuration as
\begin{equation}
\frac{d\theta}{d\tau} \simeq -\frac{1}{3}\theta^2 - \frac{1}{4}R.
\label{IsotropicRaychaudhuri}
\end{equation}
Eq. \eqref{IsotropicRaychaudhuri} shows that, once the invariant degeneracies are satisfied, the focusing of timelike geodesics is governed by the scalar curvature. Therefore, all anisotropic channels associated with shear and trace-free Ricci distortions are dynamically eliminated, yielding an isotropic approach to the singular limit. 

\section{Conclusion}
In this work we have demonstrated that polynomial degeneracies of Ricci invariants impose generic constraints on the admissibility of spacetime singularities in modified theories of gravity. These constraints arise solely from the algebraic structure of Riemannian geometry and do not depend on assumptions regarding equations of state, energy conditions or a specific solution. For spherically symmetric configurations, the degeneracies enforce an invariant suppression of the trace-free Ricci tensor at singular roots. This condition translates directly into the vanishing of effective anisotropy and removes the Ricci-sourced production of shear. In the classes of modified gravity examined in this work, this geometric selection mechanism acquires a concrete dynamical interpretation. In metric $f(R)$ gravity, the effective stress-energy tensor contains higher-derivative curvature terms which generically induce anisotropic stresses even when the physical matter sector is isotropic. Unless dynamically compensated by matter anisotropy or suppressed by the functional form of $f(R)$, they can act as persistent sources for shear and obstruct the invariant condition $S_{\mu\nu}^{\rm eff} \to 0$ required by the syzygies. A similar conclusion follows in the Brans-Dicke theories, where spatial gradients of the scalar field generate effective anisotropy. In such cases, the syzygy constraints restrict admissible singular configurations to sectors where the scalar field is homogeneous or the non-minimal coupling dynamically screens scalar-induced anisotropy. In generalized Brans-Dicke and Chameleon models, this requirement translates into constraints on the coupling functions, thereby singling out a narrow functional class compatible with singular collapse. The resulting differential constraints on $\omega(\phi)$ or $f(\phi)$ arise purely from invariant degeneracy and are independent of phenomenological or observational considerations. \\

Polynomial degeneracies therefore do not merely constrain admissible matter models, but act as invariant regulators of geodesic focusing. They control the generation of effective shear and enables a unified geometric formalism to understand singularities across GR and its extensions. One can say that spacetime singularities are not the generic outcomes of gravitational collapse in modified gravity, but occur only along algebraic branches permitted by the underlying Riemannian geometry. In this sense, classical gravity exhibits a form of geometric pre-selection of singular end states, encoded at the level of curvature invariants prior to any quantum completion. One may consider extending this framework into rotating or non-spherically symmetric spacetimes, examining Weyl-invariant syzygies and exploring whether similar invariant selection rules persist in Palatini or metric-affine formulations of gravity.

\section*{Acknowledgement}

The author acknowledges the IUCAA for providing facility and support under the visiting associateship program. Acknowledgment is also given to Vellore Institute of Technology for the financial support through its Seed Grant (No. SG20230027), the year 2023.

\end{document}